\documentclass[a4paper, 12pt]{article}

\usepackage{axodraw}
\usepackage[english]{babel}
\usepackage{cite}
\usepackage{a4wide}
\usepackage{amsmath}
\usepackage[charter,cal=cmcal,greekuppercase=italicized]{mathdesign}
\usepackage{dsfont}
\usepackage[T1]{fontenc}
\usepackage{slashed}
\usepackage{setspace}
\usepackage{graphicx}
\usepackage{xcolor}
\usepackage{soul}
\usepackage[final,          % override "draft" which means "do nothing"
            colorlinks,     % rather than outlining them in boxes
            linkcolor=purple, % override truly awful colour choices
            citecolor=teal, %   (ditto)
            urlcolor=blue  %   (ditto)
            ]{hyperref}

\makeatletter
\DeclareRobustCommand*{\bfseries}{%
  \not@math@alphabet\bfseries\mathbf
  \fontseries\bfdefault\selectfont
  \boldmath
}
\makeatother

\newcommand{\sq}{\ensuremath{\mathrm{s}}}
\newcommand{\cq}{\ensuremath{\mathrm{c}}}
\newcommand{\dq}{\ensuremath{\mathrm{d}}}
\newcommand{\uq}{\ensuremath{\mathrm{u}}}

\newcommand{\dd}{\ensuremath{\operatorname{d}}}

\newcommand{\diag}{\operatorname{diag}}

\newcommand{\SU}{\ensuremath{\mathrm{SU}}}
\newcommand{\U}{\ensuremath{\mathrm{U}}}

\newcommand{\CP}{\ensuremath{CP}}

\renewcommand{\Re}{\ensuremath{\operatorname{Re}}}
\renewcommand{\Im}{\ensuremath{\operatorname{Im}}}
\newcommand{\Mat}[1]{\ensuremath{\boldsymbol{#1}}}

\newcommand{\ie}{i.\,e.~}

\newcommand{\eg}{e.\,g.~}

\usepackage[font=small]{caption}

\begin{document}

\onehalfspacing

\begin{titlepage}

\vspace*{-15mm}
\begin{flushright}
DESY 16-084
\end{flushright}
\vspace*{0.7cm}

\begin{center} {
\bfseries\LARGE
A bottom-up approach to the strong \CP{} problem
}\\[8mm]
J.~L.~Diaz-Cruz$^{a,}$
\footnote{\texttt{jldiaz@fcfm.buap.mx}},\quad
W.~G.~Hollik$^{b,}$
\footnote{\texttt{w.hollik@desy.de}},\quad
and\quad
U.~J.~Saldana-Salazar$\,^{a,}$
\footnote{\texttt{usaldana@fcfm.buap.mx}}
\\[1mm]
\end{center}
\vspace*{0.50cm}
\centerline{$^a$ \itshape
Facultad de Ciencias F\'isico-Matem\'aticas,
Benem\'erita Universidad Aut\'onoma de Puebla,}
\centerline{\itshape C.P. 72570, Puebla, Pue., M\'exico.}
\centerline{$^b$ \itshape
Deutsches Elektronen-Synchrotron (DESY),}
\centerline{\itshape
Notkestra\ss{}e 85, D-22607 Hamburg, Germany.}
\vspace*{1.20cm}

\begin{abstract}
  The strong \CP{} problem is one of many puzzles in the theoretical
  description of elementary particle physics that still lacks an
  explanation. While \emph{top-down} solutions to that problem usually
  comprise new symmetries or fields or both, we want to present a
  rather \emph{bottom-up} perspective. The main problem seems to be
  how to achieve small \CP{} violation in the strong interactions
  despite large \CP{} violation in weak interactions. Observation of
  \CP{} violation is exclusively through the Higgs--Yukawa
  interactions. In this paper, we show that with minimal assumptions
  on the structure of mass (Yukawa) matrices they do not contribute to
  the strong \CP{} problem and thus we can provide a pathway to a
  solution of the strong \CP{} problem within the structures of the
  Standard Model and no extension at the electroweak scale is
  needed. However, to address the flavor puzzle, models based on
  minimal \(\SU(3)\) flavor groups leading to the proposed flavor
  matrices are favored. Though we refrain from an explicit a UV
  completion of the Standard Model, we provide a simple requirement
  those models should have to intrinsically not show a strong \CP{}
  problem.
\end{abstract}

\end{titlepage}

\setcounter{footnote}{0}

\section{Introduction}
%Problems are the nourishment of theoretical physics.
The Standard Model (SM) is known to be an incomplete model full of
unresolved problems. Among the issues of the SM that still wait to be
solved, the strong \CP{} problem appears to be a very central one as
it resides in the interplay of non-perturbative effects in Quantum
Chromodynamics (QCD) and \CP{} violation (CPV) in weak
interactions. Curiously, the majority of present day solutions to many
of the problems in high energy physics obey the tendency of going
beyond the SM introducing new physics at a higher scale; the strong
\CP{} problem seems to follow this tendency. However, here we take a
different philosophy and carefully scrutinize the available structures
of the SM, offering an alternative approach of the problem. Our point
of view might be defined as pragmatic, rather bottom-up, as we only
study the mass matrices along with the bi-unitary transformations
diagonalizing them. Starting from the SM flavor structures, we are
going to present guidelines for model builders to fell the strong
\CP{} problem.
%This study reveals the structure associated with \CP{} violation
%in strong interactions.
Before moving to the details of our treatment let us first briefly
summarize what the strong \CP{} problem is. For a comprehensive review
of this problem, see for instance~\cite{Mohapatra:1986uf}.

The \(\theta_\text{QCD}\) parameter of QCD parametrizes the
non-equivalence of possible QCD vacua as for non-abelian gauge fields
there can be non-vanishing winding numbers defined as
\begin{equation}\label{eq:winding}
n = \frac{g^2}{32\pi^2} \int\dd^4 x F^a_{\mu\nu} {\tilde F}^{a\, \mu\nu},
\end{equation}
leading therefore to an effective action
\begin{equation}\label{eq:effact}
S_\text{eff} = \int\dd^4x\mathcal{L} + in\theta_\text{QCD}.
\end{equation}
The axial anomaly introduces via
\begin{equation}\label{eq:axialcurrent}
\partial^\mu j^5_\mu = \frac{g^2}{16\pi^2} F_{\mu\nu}^a {\tilde F}^{a\,\mu\nu},
\end{equation}
effectively a change in \(S_\text{eff}\) by rotations of the quark
fields with \(\exp(-i\gamma_5 \frac{\theta_q}{2})\) that shifts the gauge
field \(\theta_\text{QCD}\) parameter as
\begin{equation}\label{eq:thetabar}
\theta_\text{QCD} \to \bar \theta = \theta_\text{QCD} + \theta_q.
\end{equation}

The same transformation affects quark mass terms as \(m \bar q_L q_R
\to e^{-i \theta_q} m \bar q_L q_R\) and may conversely be used to
trace \CP{} violating effects stemming from the masses. Due to this
property, we may identify the physical remaining phase after such
rephasing with the axial phase transformation and be left with the
well-known combination \(\bar\theta = \theta_\text{QCD} +
\theta_\text{QFD}\), where the quark flavor dynamical contribution is
given by
\begin{equation}\label{eq:thetaQFD}
  \theta_\text{QFD} = \arg\det\big(\Mat{M}_\uq\big)
  + \arg\det\big(\Mat{M}_\dq\big)
  = \arg\det\big(\Mat{M}_\uq \Mat{M}_\dq\big).
\end{equation}

The parameter \(\bar\theta\) violates \CP{} and induces an electric
dipole moment for the neutron, so bounds are roughly $\bar{\theta} <
10^{-10}$~\cite{Baker:2006ts}. Such a huge cancellation between those
two contributions in Eq.~\eqref{eq:thetabar} is to be seen as a
fine-tuning problem as they are conceptually independent. The strong
\CP{} problem now manifests itself in the question why \(\bar\theta\)
is so small although CPV in weak interactions has been found to be
rather large (large, of course, compared to \(\bar\theta\), not on
absolute grounds). Even though both contributions, the pure gauge
\(\theta_\text{QCD}\) and the quark flavored \(\theta_\text{QFD}\),
cannot be treated separately because a chiral phase shift in the
quarks always reintroduces a genuine \(\theta_\text{QFD}\)-term
according to Eq.~\eqref{eq:axialcurrent}, we want to disaggregate
especially the \(\theta_{\text{QFD}}\)-term on a flavor physics
groundwork. The viability of the approach is reflected in the fact
that one can always find a basis, in which either
\(\theta_{\text{QCD}}\) or \(\theta_{\text{QFD}}\)
vanishes~\cite{Cheng:1987gp}.
% The art of finding a basis in which both
%contributions vanish \emph{simultaneously} will be promoted in this
%paper.

Popular solutions to this problem are besides the possibility of
having one massless quark (typically the \(u\)-quark but the same
holds for a massless \(d\)-quark), the introduction of at least one
new symmetry (like an axial \(\U(1)\) or Peccei--Quinn
\cite{Peccei:1977hh} symmetry) that gets spontaneously (or
softly~\cite{Mohapatra:1982ib, Mohapatra:1983xw}) broken and comprises
a light pseudo Nambu--Goldstone boson, the
axion~\cite{Weinberg:1977ma, Wilczek:1977pj}.\footnote{Axions and
  axion-like particles (ALPs) have a very rich phenomenology,
  summarized \eg in~\cite{Kuster:2008zz}, with an ongoing experimental
  effort to detect them (as the ALPS experiment at
  DESY~\cite{Ehret:2007cm, Ehret:2010mh} and future facilities like
  ALPS-II or SHiP at CERN~\cite{Alekhin:2015byh}).} A third way to
compass the problem is via mechanisms worked out by
Nelson~\cite{Nelson:1983zb} and Barr~\cite{Barr:1984qx, Barr:1984fh},
for a recent review see~\cite{Dine:2015jga}. The main requirement for
this mechanism is a vanishing \(\arg\det\big(\Mat{M}_q\big)\) and a
way to spontaneously break \CP{} in the context of Grand Unified
Theories, alternatively spontaneously~\cite{Barr:1991qx} or softly
broken parity~\cite{Babu:1989rb} (or a combination of all of
them~\cite{Kuchimanchi:2010xs, Kuchimanchi:2012xb}). This proposal is
noteworthy in the sense that it provides a solution to the strong
\CP{} problem with no low energy consequence; unlike the axion and
$m_u=0$ solutions~\cite{Dine:2015jga}. It was also shown in a certain
kind of toy model that explicit soft \CP-violation in the Higgs
potential with two complex scalars leads spontaneously to an explicit
\CP-violating effect in the quarks mass matrices and still keeps
\(\det \left( \Mat{M}_\uq \Mat{M}_\dq \right)\) real at the
tree-level~\cite{Georgi:1978xz}. In this spirit, the study of the SM
structure as an effective theory is sufficient to circumvent the
original problem as one is left with the open question about the
origin of the Yukawa interactions. Another approach with spontaneous
CPV is the one involving discrete flavor
symmetries~\cite{Antusch:2013rla}. For last, generalized
$P$-invariance in left-right symmetric theories can also provide
valuable methods on computing approximately $\bar{\theta}$ through the
corresponding right-handed quark mixing
matrix~\cite{Maiezza:2014ala,Senjanovic:2014pva,Senjanovic:2015yea},
while supersymmetry helps to protect \(\bar\theta =
0\)~\cite{Barr:1996wx, Hiller:2001qg}.

Interpreting \(\theta_\text{QCD}\) as a Lagrangian parameter, it is
the only parity violating term in the QCD Lagrangian (and because
charge conjugation is conserved, the \(\theta_\text{QCD}\)-term
explicitly violates \CP). In the bottom-up approach, we take a
vanishing \(\theta_\text{QCD}\) for granted and unlike the
Nelson--Barr approach stay at first ignorant about possible symmetries
and physics at higher energy scales. Whether or not some variant of
\(P\) or \CP{} has to be employed as symmetry of nature depends on the
specific realization.  Instead, we pursue the option that CPV shall
only arise from the Yukawa interactions alone, not even from the Higgs
vacuum expectation values that multiply the Yukawa couplings. We do
not require necessarily spontaneous CPV but allow in principle for
explicit breaking in the (effective) Yukawa couplings. Imposing global
\CP-invariance of the QCD gauge interactions (though parity is enough)
suffices for our main argument. We treat the Yukawa Lagrangian as an
effective Lagrangian hiding the UV completion in the dimensionless
Yukawa matrices.  In that way, we obtain \(\theta_\text{QCD} = 0\) and
hence reduce the problem to understand why \(\arg\det\big(\Mat{M}_\uq
\Mat{M}_\dq\big)\) is such a small number (or why it should exactly
vanish).

In the course of this paper, we accordingly suppose
$\theta_\text{QCD}=0$ and show how \(\arg\det\big(\Mat{M}_q\big)\)
vanishes by imposing a minimal constraint on certain flavor phases and
still providing sufficiently large weak CPV.\footnote{We call
  \emph{flavor phases} the ones that appear both in the Yukawa
  matrices in a certain weak basis and remain, after moving to the
  mass basis, in the Jarlskog invariant. Of course, within the minimal
  SM only one physical will remain (that would be a linear combination
  of all the initial ones). But more physical phases could remain in
  extensions of the SM such as those including a fourth family and $N$
  extra Higgs doublets.}  We shall argue that the physical CPV in the
weak interactions is unrelated to any other phases appearing in
\(\arg\det\big(\Mat{M}_q\big)\) and may only give a small finite
contribution at higher orders as \(\theta_\text{QCD} = 0\) at
tree-level~\cite{Ellis:1978hq}. In the following, we give an explicit
example of symmetry structures of mass matrices based on special
unitary transformations that have the desired property and start by
finding the minimal assumptions for the mass matrices to fulfill
that. These assumptions easily find their way in any extension of the
SM that generate Yukawa couplings dynamically, either by spontaneous
breaking of the underlying flavor symmetry or by the moduli fields of
string theory \cite{Konstandin}. The generalization to an arbitrary
number of quark families illustrates the universal validity of our
idea.

Wrapping up our philosophy in order to clarify the new approach, we
want to point out that a deliberate solution to the strong \CP{}
problem does not require a distinct statement about fundamental
CPV. First of all, \CP{}-invariance of the gauge interactions is
sufficient, as we will show, to circumvent one aspect of the strong
\CP{} problem. Second, \CP{} may be violated explicitly (or
spontaneously) in the Yukawa interactions of SM fermions to the Higgs
scalar as these interactions are the less understood in the context of
the SM and do not necessarily have to respect \CP. Third, we shall
identify types of fermion mass matrices that automatically cancel out
the undesired contribution to the \(\bar\theta\)-parameter and thus
solve the strong \CP{} problem without the need of additional degrees
of freedom in the theory as new fields assisting in this
process. Finally, we stay open towards the remaining solution of the
flavor puzzle in the SM, namely the question why there are three
families of fermions and why they behave and mix as they do. However,
we give an explicit realization of a parametrization (not yet a model)
that helps to explain the mixing angles fully in terms of mass ratios
and additionally provides the weak \CP-phase of the SM in the same
form.

This paper is organized as follows. In Section
\ref{sec:weakandstrong}, we disentangle the origins of strong and weak
CPV despite the fact that both of them can be expressed in terms of
the same quark mass matrices. In Section \ref{sec:weakCPV}, we write
the Kobayashi--Maskawa phase in terms of quark mass ratios as it
follows from~\cite{Hollik:2014jda}, which is seen to be independent
from the previous considerations. In Section \ref{sec:strongCPV}, we
study general consequences of the conditions provided in this work and
how they are related to concrete models. Finally, we conclude in
Section \ref{sec:concl}.

\section{Disentangling weak and strong CPV}
\label{sec:weakandstrong}
A complete knowledge of the quark mass matrices, $\Mat{M}_\uq$ and
$\Mat{M}_\dq$, tackles down the flavor puzzle in the SM and finally
evades the strong \CP{} problem (since the quark flavor
contribution to \(\bar\theta\) can be absorbed in the masses and
\(\theta_\text{QCD} = 0\)). Strong and weak CPV, \(\theta_\text{QFD}\)
and the rephasing invariant $J_q$, respectively, are of different
origins on one hand. And on the other hand sufficiently large CPV in
the weak sector\footnote{The expression ``large'' does not have to be
  correlated to a numerically large value of the only \CP-violating
  phase in weak interactions, since, depending on the chosen
  parametrization, this phase can be rather small though the overall
  \CP-violation appears to be actually large
  (see~\cite{Gerard,Gerard:2012ft}). The rephasing invariant \(J_q\)
  does not depend on the particular parametrization and serves as a
  better measure of comparison.}  therefore does not necessarily imply
strong CPV despite the fact that both reside in the same mass matrices
\begin{itemize}
\item weak CPV: the Jarlskog invariant\\ $J_q \sim
  \Im\left[{{\text{det}}([\Mat{M}_\uq \Mat{M}_\uq^\dagger,\Mat{M}_\dq
  \Mat{M}_\dq^\dagger])}\right]$, see~\cite{Jarlskog:1985ht},
\item strong CPV: the \(\theta_{\text{QFD}}\)-term from above\\
\(\theta_{\text{QFD}} \equiv
  \arg\det\big(\Mat{M}_\uq\Mat{M}_\dq\big)\).
\end{itemize}
It is not only the functional dependence on the mass matrices (and
thus relevant phases) that is apparently different for weak and strong
CPV but rather their individual transformation properties of parity
and charge conjugation~\cite{Gerard}. While both objects are \CP-odd,
\(\theta_\text{QCD}\) transforms even under \(C\) and odd under \(P\)
whereas \(J_q\) behaves the other way round,
see~\cite{Gerard:2008nc}. For ``complete knowledge'' of mass matrices,
it is adequate to set up each matrix in terms of known parameters,
even though a full theory of flavor is still lacking. By the freedom
of relying on an effective description of the mass matrices, we
encompass the strong \CP{} problem by an ansatz to understand the
flavor puzzle.

In the following, we will assume that all quark masses are different
from zero, as suggested by lattice
calculations~\cite{Aoki:2013ldr}. Our minimal requirement for the mass
matrices follows very obviously from the usual Singular Value
Decomposition\footnote{A similar conclusion can be drawn by virtue of
  the Polar Decomposition, \(\Mat{M}_q = \Mat{H}_q
  \Mat{U}_q\)~\cite{Gerard, Gerard:2008nc}, with \(\Mat{H}_q\) a
  Hermitian and \(\Mat{U}_q\) a unitary matrix (note that \(\Mat{U}_q
  \neq \Mat{R}_q\) in the Singular Value Decomposition).}
\begin{equation}\label{eq:SVD}
\Mat{M}_q = \Mat{L}_q^\dag \Mat{\Sigma}_q \Mat{R}_q,
\end{equation}
with \(\Mat{L}_q\) and \(\Mat{R}_q\) unitary transformations and
\(\Mat{\Sigma}_q = \diag(m_{q_1}, m_{q_2}, m_{q_3}) \) a positive
diagonal matrix, \(m_{q_i} > 0\). Consequently, we find
\begin{equation}
\begin{aligned}
  \arg\det\big(\Mat{M}_\uq \Mat{M}_\dq\big) &=
  \arg\det\big( \Mat{L}_\uq^\dag \Mat{\Sigma}_\uq \Mat{R}_\uq
  \Mat{L}_\dq^\dag \Mat{\Sigma}_\dq \Mat{R}_\dq \big) \\ &=
  \arg\left(\det\Mat{L}_\uq^\dag \det\Mat{R}_\uq \det\Mat{L}_\dq^\dag
    \det\Mat{R}_\dq
\right),
\end{aligned}
\end{equation}
after using the well-known property of the determinant,
\(\det(\Mat{A}\Mat{B}) = \det\Mat{A} \det\Mat{B}\), with the \(\arg\)s
of the diagonal matrices vanishing (as they are real and
positive). The decomposition of Eq.~\eqref{eq:SVD} is completely
arbitrary in the sense that different choices of \(\Mat{L}_q\) and
\(\Mat{R}_q\) lead to different flavor representations of the mass
matrices and additionally the particular choice of \(\Mat{L}_\uq =
\Mat{L}_\dq\) leaves the weak interaction basis invariant. In case,
one desires to build mass matrices \(\Mat{M}_q\) on a certain family
of flavor symmetries, this fixes the allowed classes of
transformations \(\Mat{L}_q\) and \(\Mat{R}_q\). We give constraints
on these transformations that can be easily verified in any proposal
of a flavor model to be inherently free of a strong \CP{} problem.
  %\footnote{Of course, there are still \emph{arbitrary} unitary
  %transformations possible that rotate the mass matrices in whatever
  %shape, \(\tilde{\Mat{M}}_\uq = \Mat{L}^\dag_Q \Mat{M}_\uq
  %\Mat{R}'_\uq\) and \(\tilde{\Mat{M}}_\dq = \Mat{L}^\dag_Q \Mat{M}_\dq
  %\Mat{R}'_\dq\). These transformations, however, if not part of the
  %allowed flavor transformations break the proposed symmetry in the mass
  %terms \(\bar Q_{\Ll,i} \left( \Mat{M}_\uq \right)_{ij} u_{\Rr,j} +
  %\bar Q_{\Ll,i} \left( \Mat{M}_\dq \right)_{ij} d_{\Rr,j}\).}

In the limit of vanishing mass matrices, the SM Lagrangian (the
kinetic terms) obeys for \(n\) generations a \(\U(n)\) symmetry for
each gauge multiplet (\ie{}left-handed quark doublets and the up- and
down-type right-handed singlets; in total \(\U(n)^3\)). The maximal
freedom to rotate these fields is parametrized hence by $\U(n)$
transformations \(\Mat{U}\) with $\det\Mat{U} = e^{i\phi}$, such that
\begin{equation}\label{eq:argdetphases}
\arg\det\big(\Mat{M}_\uq \Mat{M}_\dq\big) = \alpha_R^{(u)} -
\alpha_L^{(u)} + \alpha_R^{(d)} - \alpha_L^{(d)},
\end{equation}
where \(\arg\det\big(\Mat{K}_q\big) = \alpha_K^{(q)}\) and \(\Mat{K}\)
either \(\Mat{L}\) or \(\Mat{R}\). Let us remark that this result is
independent of \(n\) and applies in the same way for any number of
fermion families, see Table \ref{table-phases}.

\begin{table}
  \caption{Complex phases contributing to either the strong or weak
    CPV phase with an arbitrary number of fermion families. Notice how
    always the same linear combination of phases appears in the strong
    \CP{} case. Through this table it becomes very apparent that the
    origin of strong CPV is completely independent from the one
    generating quark mixing or weak CPV.}
\label{table-phases}
\centering
\begin{tabular}{ccc}
\hline \hline
Number of fermion families & $\theta_{\text{QFD}}$ & $\delta_{\CP}^{\text{weak}}$  \\
\hline \hline
1  & $\alpha_R^{(u)} -
\alpha_L^{(u)} + \alpha_R^{(d)} - \alpha_L^{(d)}$ & 0 \\
\hline
2  & $\alpha_R^{(u)} -
\alpha_L^{(u)} + \alpha_R^{(d)} - \alpha_L^{(d)}$  & 0 \\
\hline
3  & $\alpha_R^{(u)} -
\alpha_L^{(u)} + \alpha_R^{(d)} - \alpha_L^{(d)}$  & $\delta^{\text{CKM}}_{\CP}$\\
\hline
4  & $\alpha_R^{(u)} -
\alpha_L^{(u)} + \alpha_R^{(d)} - \alpha_L^{(d)}$  & $\delta^{\text{CKM}}_{\CP}, \omega_{\CP}, \omega'_{\CP}$ \\
\hline
$n$  & $\alpha_R^{(u)} -
\alpha_L^{(u)} + \alpha_R^{(d)} - \alpha_L^{(d)}$  & $\delta^{\text{CKM}}_{\CP}, \; \omega_{\CP}^k \quad (k=1,2,\cdots,\frac{n(n-3)}{2} )$ \\
\hline
\hline
\end{tabular}
\end{table}
%(n-1)(n-2)/2

We emphasize that strong and weak CPV, $\theta_{\text{QFD}}$ and
$J_q$, are of different origin and thus can be treated
independently. If the left and right unitary rotations \(\Mat{L}_q\)
and \(\Mat{R}_q\), respectively, were either \emph{special} unitary
rotations or the same (as for Hermitian mass matrices),
Eq.~\eqref{eq:argdetphases} would vanish trivially and still allow for
a non-vanishing Jarlskog invariant. [Recall that unitary
  transformations are equal to special unitary transformations times a
  global phase, \(\U(n) =\SU(n)\otimes \U(1)/Z_n \) \cite{Salcedo}.]
Hence, the global phases give rise to strong CPV and consequently
$\theta_{\text{QFD}}$ is sensitive to these global phases only and
insensitive to the complex structure of the underlying \(\SU(3)\)
transformations responsible for flavor mixing phenomena. Weak CPV on
the other hand has exactly the opposite relation: it is sensitive to
the complex nature of the special bi-unitary transformations (which
give rise to flavor mixing in weak interactions) and insensitive to
global phases (a property exactly expressed by Jarlskog invariant
which is known to be rephasing invariant). Even in the absence of a
global phase that violates \CP{} strongly, we can have sufficiently
large weak CPV: the existence or non-existence of strong CPV is
completely unrelated to the existence or non-existence of weak CPV as
can be seen from the two family case which has no weak CPV while
showing, in principle, strong CPV. Conversely, comparison of the
expressions of \(J_q\) and \(\theta_{\text{QFD}}\) as listed above
already shows that the phase difference of \(\Mat{M}_\uq\) and
\(\Mat{M}_\dq\) responsible for \(\theta_{\text{QFD}} \neq 0\) drops
out in the expression of \(J_q\). Note that the existence of these
global \(\U(1)\)-phases and the invariance of the SM field content
under a particular combination of such rephasings is just the
conservation of baryon number, which, however, is accidental.

We have shown that the strong CPV parameter ($\theta_\text{QFD}$) can
be treated independently from the weak one, which already on its own
is quite distinct from other approaches. However, to identify how the
above conditions feed into a viable UV complete model one needs to go
farther.  In this regard, it helps to recognize certain benchmark
scenarios that have $\theta_\text{QFD} = \alpha_R^{(u)} -
\alpha_L^{(u)} + \alpha_R^{(d)} - \alpha_L^{(d)} = 0$.  The different
ways to achieve this goal are sketched in Table \ref{table-cases},
note that in principle one should be able to smoothly interpolate
between those scenarios. These conditions should be seen as applied to
a more fundamental theory where $\theta_\text{QCD} = 0$ and which when
taken to lower energies one delivers the SM. In this sense, the Yukawa
part of the SM Lagrangian could be treated as an effective
Lagrangian. Case Ia considers a CP-invariant Lagrangian where there
are no phases; in order to get weak CPV one must either spontaneously
or explicitly break it. Case Ib refers to a flavor theory employing an
$\SU(3)$ symmetry group or subgroup. Case II embraces those Left-Right
(LR) symmetric theories where parity is conserved. Cases III and IV,
are other examples with no explicit model present in which the sum of
global phases could be canceled.

\begin{table}
  \caption{Different cases for which there is no contribution in the
    strong CP phase, $\bar{\theta}$, stemming from the global phases,
    $\theta_{\text{QFD}}$.} \label{table-cases} \centering
\begin{tabular}{ccccccc}
\hline \hline
Case & $\alpha_L^{(u)}$ & $\alpha_L^{(d)}$ & $\alpha_R^{(u)}$ & $\alpha_R^{(d)}$ & Condition on the Yukawa couplings & Weak CPV  \\
\hline \hline
Ia  & $0$ & $0$ & $0$ & $0$ & $CP$-invariance & No \\
\hline
Ib  & $0$ & $0$ & $0$ & $0$ & ${\cal G} \subseteq \SU(3)$-invariance & Yes \\
\hline
II  & $\alpha_u$ & $\alpha_d$ & $\alpha_u$ & $\alpha_d$ & $P$-invariance & Yes\\
\hline
III & $\alpha_L$ & $-\alpha_L$ & $\alpha_R$ & $-\alpha_R$ & Unknown & Yes\\
\hline
IV & $\alpha$ & $\beta$ & $\beta$ & $\alpha$ & Unknown & Yes\\
\hline
V & \(\alpha_L\) & \(\alpha_L\) & \(x\) & \(2 \alpha_L - x\) &
  \(\SU(2)_L\) gauge & ? \\
\hline
\hline
\end{tabular}
\end{table}

Therefore, within this context, the vanishing
\(\arg\det\big(\Mat{M}_\uq\Mat{M}_\dq\big)\) is automatic, contrary to
the common folklore. Instead, certain conditions, that are summarized
in Table~\ref{table-cases}, could be taken as forced by symmetry
reasons.  Our approach to solve the strong \CP{} problem reduces
essentially to explain why \(\alpha_L^{(q)} = \alpha_R^{(q)}\)
(similar to Case II in Table \ref{table-cases} without explicit
\(P\)-invariance) while, simultaneously, explain the observed amount
of weak CPV, $\delta_{\CP}^{\text{CKM}} = (1.19 \pm 0.15 )\text{
  rad}$, see~\cite{Agashe:2014kda}.

Before we move to an explicit realization of our findings, let us
briefly summarize what we have so far: even if all quarks are massive,
we have no strong \CP{} problem without imposing any new
symmetries. The requirement \(\arg\det\big(\Mat{M}_\uq
\Mat{M}_\dq\big) = 0\) can be achieved by all of the benchmark cases
of Table \ref{table-cases} and any interpolation between them. For
example, Case I or II, ensures \(\alpha_L^{(q)}\) and
\(\alpha_R^{(q)}\) for \(q = u,d\) to be zero or equal, respectively;
the minimal way for the former scenario would be to propose either
\CP-invariance or \(\SU(3)\) transformations for the diagonalization
of the mass matrices which conversely means that any flavor model
based on \(\SU(3)\) transformations gives a solution to the strong
\CP{} problem \cite{Grimus:2010ak, Grimus:2011fk}. Finally, we can
still have (arbitrarily large) CPV in weak interactions as this is
unrelated to strong CPV. The main task is somewhat to reduce the
arbitrariness in complex phases that are generally allowed for the
mass matrices and give a restrictive prescription for weak CPV. It is
comparably simple to define generic \CP-violating textures of mass
matrices that have no strong \CP-phase and still allow for a CKM-phase
according to a mismatch between phases in up and down type mass
matrices~\cite{Holdom:1999ny}. Our approach, however, is still even
more generic as we do not rely on a certain flavor basis in which the
phases are apparent and stay rather basis and model independent.

\section{A suggestive way of calculating weak CPV}
\label{sec:weakCPV}

In the previous section, we have stated that the $\SU(3)$ symmetry
transformations in family space, acting in the left and right handed
fields of up-quark and down-quark types, are enough to deliver any
amount of weak CPV independently of having previously eliminated the
combination of $U(1)$ global phases which give rise to the strong
\CP{} phase, $\theta_\text{QFD}$. Now, we want to outline in a
recently proposed mixing parametrization~\cite{Hollik:2014jda}, how to
represent the weak CP phase. This parametrization relates the entries
of the Cabibbo--Kobayashi--Maskawa (CKM) matrix to functions of the
quark mass ratios and thus the remaining CP phase can be written as
function of those.

A very famous expression of a mixing angle as a function of a mass
ratio was provided by the well-known Gatto--Sartori--Tonin (GST)
relation, \(\tan\theta_C \approx \sqrt{m_\dq / m_\sq}\), for the
Cabibbo angle \(\theta_C\)~\cite{Gatto:1968ss}. Based on this finding,
a parametrization of the fermion mixing matrices was proposed that
only uses the mass ratios as input~\cite{Hollik:2014jda}. Besides the
phenomenological observation \(m_i \ll m_j\) for \(i<j\) with masses
of the \(i\)-th and \(j\)-th generation, a crucial assumption behind
this parametrization is that the Euler rotations can be individually
expressed by \(\tan\theta_{ij} = \sqrt{m_i / m_j}\). Likewise,
symmetrical structures in the mass matrices have been detected that
lead to exactly this kind of mixing
matrices~\cite{Saldana-Salazar:2015raa}. In that view, the final quark
mixing matrix can be decomposed into a chain of successive two-family
rotations where each planar \(\SU(2)\) rotation can then be written as
\begin{equation}\label{eq:2by2}
\Mat{U}'_{ij}(\mu_{ij},\delta_{ij}) = \begin{pmatrix}
\frac{1}{\sqrt{1+\mu_{ij}}} & \sqrt{\frac{\mu_{ij}}{1+\mu_{ij}}}
e^{-i\delta_{ij}} \\ - \sqrt{\frac{\mu_{ij}}{1+\mu_{ij}}}
e^{i\delta_{ij}} & \frac{1}{\sqrt{1+\mu_{ij}}} \end{pmatrix},
\end{equation}
with \(\mu_{ij} = m_i / m_j\) and an \emph{a priori} arbitrary complex
phase \(\delta_{ij} \in [0, 2\pi)\). We identify \(\sin\theta_{ij} =
\sqrt{\frac{\mu_{ij}}{1+\mu_{ij}}}\) and \(\cos\theta_{ij} =
\frac{1}{\sqrt{1+\mu_{ij}}}\).  For example, the full \(\SU(3)\)
transformation for the the 2-3 sub-sector is then given by
\begin{equation}
  \Mat{U}'_{23} \left(\frac{m_2}{m_3},\delta_{23} \right) =
  \begin{pmatrix}
    1 & 0 & 0 \\
    0 &\frac{1}{\sqrt{1+m_2/m_3}} & \sqrt{\frac{m_2 /m_3}{1+m_2/m_3}}
    e^{-i\delta_{23}} \\
    0 & - \sqrt{\frac{m_2/m_3}{1+m_2/m_3}} e^{i\delta_{23}} &
    \frac{1}{\sqrt{1+m_2/m_3}}
  \end{pmatrix}.
\end{equation}

Now, defining the CKM-matrix as
\begin{equation}\label{eq:defCKM}
\Mat{V}_\text{CKM} \equiv \Mat{L}_\uq \Mat{L}_\dq^\dag,
\end{equation}
with the \(\U(3)\) transformations \(\Mat{L}_{\uq,\dq}\) defined via
Eq.~\eqref{eq:SVD}, we have in the formulation
of~\cite{Hollik:2014jda} four mass ratios entering the game and six
phases from which three can be removed by choosing the up-type mass
matrix real.\footnote{This rephasing should not introduce a new strong
  \CP-phase as we only shuffle complex entries from \(\Mat{M}_\uq\) to
  \(\Mat{M}_\dq\). Moreover, any global phase does not play a role for
  weak \CP-violation as the relevant objects are the left-Hermitian
  products \(\Mat{M}_{\uq,\dq} \Mat{M}_{\uq,\dq}^\dag\).} From the
remaining three, only \emph{one maximally \CP-violating phase} sitting
in the 1-2 rotation is needed to fully reproduce the CKM-phase,
details may be found in~\cite{Hollik:2014jda}. It was also pointed out
in Ref.~\cite{Antusch:2009hq} that the same follows for certain 1-3
texture zero mass matrices. The approach of~\cite{Hollik:2014jda} is
however more general as it does not rely on specific texture zeros but
merely on symmetrical structures \`a la
Ref.~\cite{Saldana-Salazar:2015raa}.

Using this mass ratios parametrization, we similarly compute the
Kobayashi--Maskawa \CP-phase in terms of mass ratios. In the standard
parametrization, the most recent global fit obtains for it
$\delta_{\text{CP}}^{\text{CKM}} = (1.19 \pm 0.15
)\;\text{rad}$~\cite{Agashe:2014kda}.  In the following, we want to
estimate the corresponding theoretical value. After imposing
individual rotations of the type~\eqref{eq:2by2}, we can finally build
up a quark mixing matrix that has non-vanishing CPV.

The procedure introduced in Ref.~\cite{Hollik:2014jda} gives a mixing
matrix which cannot be directly compared to the conventional
parametrization. In order to do that, we first need to rephase both
the up and down type quark fields
\begin{eqnarray}
  \tilde{\Mat{V}}_\text{CKM} = \Mat{\chi}_\uq \Mat{V}_\text{CKM} \Mat{\chi}_\dq^\dagger,
\end{eqnarray}
in such a way that we are able to produce the following structure
\begin{eqnarray}
  \tilde{\Mat{V}}_\text{CKM} \sim \begin{pmatrix}
    \Re & \Re & {\cal C} \\
    {\cal C} & {\cal C} & \Re  \\
    {\cal C} & {\cal C} & \Re
  \end{pmatrix},
\end{eqnarray}
where $\Mat{\chi}_q = \diag(e^{i \phi_q},1,1)$ and $\Re$ and ${\cal
  C}$ mean real and complex entries. After rephasing, we get the
following expression for the Kobayashi--Maskawa \CP-phase
\begin{eqnarray}
  \delta^{\text{q}}_{\text{CP}} \approx
  \arctan \left[ \sqrt{
      \frac{{\mu}_{\dq\sq}(1+{\mu}_{\dq\sq})}{{\mu}_{\uq\cq}(1+{\mu}_{\uq\cq})}}
  \right] \approx (1.38 \pm 0.10) \;\text{rad},
\end{eqnarray}
which after insertion of the values of the quark mass ratios,
\(\mu_{\dq\sq} = m_\dq / m_\sq = 0.051\pm 0.001\) and \(\mu_{\uq\cq} =
m_\uq / m_\cq = 0.0021\pm 0.0001\), we find it to be in agreement to
the experimental value, \(\delta_\text{CP}^\text{CKM} = (1.19 \pm 0.15
)\).\footnote{The quark masses have been treated as running
  \(\overline{\text{MS}}\) masses evaluated at the weak scale (\(Q^2 =
  M_Z^2\)), numbers are taken from App.~A in
  Ref.~\cite{Hollik:2014jda}.} Notice how when the decoupling limit,
$m_{b,t} \rightarrow \infty$, is considered the \CP-phase does not go
to zero as one would expect it in other parametrizations
\cite{Gerard:2012ft}. Nevertheless, there is no inconsistency in this
result as simultaneously the magnitudes of the mixing matrix elements
vanish, \(\left|\Mat{V}^\text{CKM}_{13} \right| \to 0\) and
\(\left|\Mat{V}^\text{CKM}_{23} \right| \to 0\) as \(m_{t,b} \to
\infty\). Conversely, the \CP{} phase gets closer to its maximum value
\(\frac{\pi}{2}\) when the ratio between the up and the charm quark
gets more suppressed, $m_u/m_c \rightarrow 0$.

Hence, we have shown that within the SM without adding new degrees of
freedom the amount of weak CPV can be calculated by means of the quark
mass ratios if one allows for a relation among CKM angles and mass
ratios.

What about higher order corrections? The impact of the weak \CP{}
phase in the CKM model on the strong \CP{} phase was first and
extensively studied in Ref.~\cite{Ellis:1978hq} where the generic
contribution was shown to be small. However, at very high (\ie{}14th)
order in perturbation theory there is an ``infinite'' contribution
which actually turns out to be rather tiny when the original
\(\theta_\text{QCD}\) parameter is renormalized to zero at around the
Planck scale. Even with some ``New Physics'' contribution (of heavy
quarks above the electroweak scale---remembering that at the time this
reference originates the mass of the top quark was expected to be well
below its today's value) there is no huge effect. Proper New Physics
contributions, however, strongly depend on the implementation of New
Physics and shall be rather tuned to avoid a strong effect on
\(\theta_\text{QCD}\) anyway. Spurious contributions at low energies
are not to be expected once the Standard CKM model has been generated
in a top-down approach.

\section{Towards a UV completion}
\label{sec:strongCPV}

After having paved the path towards a UV complete theory of flavor
without strong \CP{} problem, we explore the possible consequences of
the conditions above to see where the path may end. Through this
reasoning, the functional dependence of the Kobayashi--Maskawa phase
on the quark mass ratios in a parametrization suggested by some of the
authors~\cite{Hollik:2014jda} helps to describe the phenomenological
properties of the true flavor dynamics realized in nature. In this
regard, we start with certain flavor matrix
structures~\cite{Saldana-Salazar:2015raa}, which have successfully
described fermion mixing~\cite{Hollik:2014jda}, and study their
connection to the strong \CP{} problem.

We explicitly refrain from providing a UV complete model extending the
field content of the SM in order not to spoil the generality of our
results. What follows shall be rather seen as matching conditions of
any theory of flavor generating mass or Yukawa matrices for the low
energy effective theory where all relevant heavy degrees of freedom
are integrated out. We leave it to either the interested reader to
construct such a model which give such conditions or to future works
of the authors. As the framework itself leading the results
in~\cite{Hollik:2014jda,Saldana-Salazar:2015raa}, is directly related
to the decomposition of $3\times3$ matrices into $2\times 2$
submatrices, we show some relations in the two-family case only,
\textit{not losing by this any generality in our treatment}.

It is outside the scope of this work to provide the full details of
the two mentioned
publications~\cite{Hollik:2014jda,Saldana-Salazar:2015raa}. Nonetheless,
here we briefly comment the essential ideas behind them which might be
incorporated in any UV complete model wishing to serve as a theory of
flavor. The first work delivers a new mixing parametrization which
applies to both quarks and leptons~\cite{Hollik:2014jda}. A
systematical procedure was built through the phenomenological
observation of hierarchical fermion masses, $m_3^2 \gg m_2^2 \gg
m_1^2$, along with a lower rank approximation theorem known as
Schimdt--Mirsky. In a similar fashion as the Wolfenstein
parametrization, where the four mixing parameters are real but still
the parametrization is complex, the four independent mass ratios of
either the quark or lepton sector are used as mixing parameters,
\begin{eqnarray}
  \Mat{V}_f = \Mat{V}_f \left( \frac{m_1^a}{m_2^a},
  \frac{m_2^a}{m_3^a}, \frac{m_1^b}{m_2^b}, \frac{m_2^b}{m_3^b}
  \right),
\end{eqnarray}
where $f= q, \ell$ is the CKM or PMNS mixing matrix, respectively, and
$a=u,\nu$ and $b=d,e$. This procedure exploits the mathematical
properties of matrices under the fact of hierarchical singular
values. There is no lost of generality in any of the involved
approximations as cautious steps are made. Two main issues, however,
are left: what symmetry or principle dictates that Yukawa matrices
should arise with the following structure,
\begin{eqnarray} \label{eq:3matrixStruc}
        \Mat{M} \sim \begin{pmatrix}
                & &  \\
                & & \\
                & & \times
        \end{pmatrix} +
        \begin{pmatrix}
                & &  \\
                & & \times \\
                & \times &
        \end{pmatrix} +
        \begin{pmatrix}
                & \times & \times \\
                \times & &   \\
                \times & &
        \end{pmatrix}?
\end{eqnarray}
and what is producing among them a hierarchy?

The second work~\cite{Saldana-Salazar:2015raa} precisely offers an
answer to the first question. The mass matrix which gets diagonalized
by a transformation of the type~\eqref{eq:2by2} can be found very
easily in the two family case and can be generalized to \(n>2\)
generations according to~\cite{Saldana-Salazar:2015raa}. In
Ref.~\cite{Saldana-Salazar:2015raa}, mass matrices are constructed in
such a way to allow the sequential diagonalization
of~\cite{Hollik:2014jda} without preference to any of the
families. The basic assumption behind this approach is that
Higgs--Yukawa interactions (or conversely mass matrices if one does
not specify the mass generating mechanism directly) are symmetric
under permutations of the fermion fields. This permutation symmetry is
then supposed to be broken stepwise as \(S_{3L} \otimes S_{3R} \to
S_{2L} \otimes S_{2R} \to S_{2A} \oplus S_{2S}\), where the last step
proceeds to a sum of anti-symmetric and symmetric permutation matrices
of two objects. This proposal can be fulfilled employing textures like
the one appearing in Eq.~\eqref{eq:3matrixStruc} which allow to study
the corresponding mixing by three different rotations in a two-family
space each.

We exemplarily study the two-family case where the mass matrix
originated in the sequential breakdown of permutation
symmetries~\cite{Saldana-Salazar:2015raa} is given in a
\emph{preferred} basis\footnote{The preferred basis corresponds to the
  mass basis for \(m_1 \to 0\).} as
\begin{eqnarray} \label{eq:FBP}
  \Mat{M} = \begin{pmatrix}
    0 & \sqrt{m_1 m_2} e^{-i\delta_m} \\
    -\sqrt{m_1 m_2} e^{i \delta_m} & m_2 - m_1
  \end{pmatrix}.
\end{eqnarray}

Now, we want to map this structure resulting in a GST relation to the
most general case of a \(2\times 2\) mass matrix. The GST relation
gives Eq.~\eqref{eq:2by2} as the corresponding unitary
transformation. A general \(\U(2)\) matrix has two more parameters
that can be expressed as an additional phase on the diagonal and an
overall phase factor,
\begin{equation}\label{eq:generic-2by2}
\Mat{U} = e^{i\phi/2} \begin{pmatrix}
\cos\theta e^{i\eta} & \sin\theta e^{-i\delta} \\ - \sin\theta
e^{i\delta} & \cos\theta e^{-i\eta} \end{pmatrix},
\end{equation}
such that \(\det\Mat{U} = e^{i\phi}\).  According to
Eq.~\eqref{eq:generic-2by2}, the relevant left rotation of a generic
mass matrix should also have the form
\begin{eqnarray}\label{eq:L2}
  \Mat{L} = e^{i \alpha_L/2} \begin{pmatrix}
    \cos \theta_L e^{i \beta_L}& \sin \theta_L e^{-i\delta_L} \\
    - \sin \theta_L e^{i\delta_L} & \cos \theta_L e^{-i \beta_L}
  \end{pmatrix}.
\end{eqnarray}
The same expression follows for the right transformation \(\Mat{R}\)
with \(L \leftrightarrow R\) in Eq.~\eqref{eq:L2} and the individual
entries of the mass matrices can be expressed via
\begin{eqnarray}
  \Mat{M} =
  e^{i\frac{\alpha_R-\alpha_L}{2}} \begin{pmatrix}
    \cos \theta_L e^{-i \beta_L}& -\sin \theta_L e^{-i\delta_L} \\
    \sin \theta_L e^{i\delta_L} & \cos \theta_L e^{i \beta_L}
  \end{pmatrix}
  \begin{pmatrix}
    m_1 & 0 \\
    0 & m_2
  \end{pmatrix}
  \begin{pmatrix}
    \cos \theta_R e^{i \beta_R}& \sin \theta_R e^{-i\delta_R} \\
    -\sin \theta_R e^{i\delta_R} & \cos \theta_R e^{-i \beta_R}
  \end{pmatrix},
\end{eqnarray}
and thus
\begin{equation}
\begin{aligned}
  M_{11} &= e^{i\frac{\alpha_R-\alpha_L}{2}} \left[ e^{-i(\beta_L - \beta_R)}m_1\cos\theta_L \cos\theta_R + e^{-i(\delta_L-\delta_R)}m_2\sin\theta_L\sin\theta_R \right], \\
  M_{12} &= e^{i\frac{\alpha_R-\alpha_L}{2}} \left[ e^{-i(\beta_L + \delta_R)} m_1 \cos\theta_L \sin\theta_R - e^{-i(\beta_R+\delta_L)}m_2 \cos\theta_R\sin\theta_L \right], \\
  M_{21} &= e^{i\frac{\alpha_R-\alpha_L}{2}} \left[  e^{i(\beta_R+\delta_L)}m_1 \cos\theta_R\sin\theta_L - e^{i(\beta_L + \delta_R)} m_2 \cos\theta_L \sin\theta_R \right], \\
  M_{22} &= e^{i\frac{\alpha_R-\alpha_L}{2}} \left[ e^{i(\beta_L -
      \beta_R)}m_2\cos\theta_L \cos\theta_R +
    e^{i(\delta_L-\delta_R)}m_1\sin\theta_L\sin\theta_R \right].
\end{aligned}
\end{equation}
Matching this set of equations to the matrix form of
Eq.~\eqref{eq:FBP} reduces the freedom of the \(\U(2)\) rotations as
the structure is dictated by the simple symmetry patterns. We find the
conditions
\begin{eqnarray}\label{eqs:conditions}
  \alpha_L = \alpha_R, \qquad
  \beta_L - \beta_R = \delta_L - \delta_R = 0, \qquad
  \theta_R = - \theta_L .
\end{eqnarray}
Identifying the impact of those relations is rather trivial in
comparison with Eq.~\eqref{eq:FBP}, as the mass matrix there clearly
exhibits no global phase (and thus \(\alpha_L = \alpha_R\)), and the
off-diagonal phase is given by \(\delta_m = \delta_{L(R)} +
\beta_{L(R)}\), as combination of the two relevant phases in the
\(\SU(2)\) rotation. Notice that, although Eq.~\eqref{eq:FBP} is not
left-right symmetric (the mass matrix is anti-Hermitian), one gets
roughly the same conditions on the left and right rotation
matrices. The last relation, \(\theta_L = -\theta_R\) follows after
commuting the phase matrices through and absorbing phases in a
redefinition of the fermion fields. This redefinition does not fully
apply to the three-family case and thus there is a remaining
\CP-violating phase in the mixing.

The constraints of Eqs.~\eqref{eqs:conditions} provide very valuable
information especially on the \emph{right-handed} rotations coded in
\(\Mat{R}\) resulting in the clear prediction that the individual
mixing angles of the right-handed sector have exactly the same
magnitude as the known left-handed (\ie{}CKM) ones. The future
detection of right-handed currents may be a razor to finally rule out
the proposed description. Note, that we do not predict right-handed
currents at all. If, however, a right-handed counterpart to the
electroweak gauge group exists, the corresponding CKM matrix cannot be
arbitrary in that description.

Fulfillment of conditions similar to~\eqref{eqs:conditions} is very
natural in left-right symmetry models. It is well-known that such
parity-invariant models offer a solution to the strong \CP{} problem
on their own without the need of an axion
solution~\cite{Babu:1989rb}. A minimal model is based on the gauge
group \(\SU(2)_L \times \SU(2)_R \times \U(1)_{B-L}\), where the left-
and right-handed fermions transform as doublets of \(\SU(2)_L\) and
\(\SU(2)_R\), respectively. Parity invariance requires the Yukawa
matrices to be Hermitian and the mass matrices are of the form
\begin{equation}\label{eq:masslinyuk}
  \Mat{M}_f= \sum_i  \Mat{Y}^{(f)}_i \langle\chi^0_i \rangle,
\end{equation}
with the vacuum expectation value (vev) \(\langle\chi_i^0 \rangle\) of
the relevant set of Higgs multiplets taking part in electroweak
symmetry breaking. Consequently, for the necessary condition on the
\emph{mass matrices}, all the vevs have to be real in order not to
spoil the Hermiticity of the Yukawa matrices. Generically, however,
such multi Higgs models easily have spontaneous \CP{} violation with
at least one complex vev. Supersymmetry helps to cure this problem,
introduces on the other hand a new strong \CP{} problem connected to
the potentially complex gluino mass~\cite{Mohapatra:1997su,
  Kuchimanchi:1995rp}.  Another avenue involves the complete doubling
of fermions and gauge group~\cite{Barr:1991qx, Chang:1993eq}, which
includes additional mirror fermions as singlets under the SM gauge
group but charged under a mirror gauge group \(\SU(2)_R \times
\U(1)_X\).  The concept of a hidden sector together with LR symmetry
applies also to radiative solutions of the flavor hierarchy
problem---and automatically complies with the conditions presented
here~\cite{Gabrielli:2016vbb}. Conversely, LR-inspired models of
flavor model building have no need for a flavored axion as recently
proposed on basis of a Froggatt--Nielsen mechanism~\cite{Ema:2016ops,
  Calibbi:2016hwq}.

We see several viable approaches to build reasonable flavor models
that are intrinsically free of the strong \CP{} problem:
\begin{itemize}
\item Multi-Higgs models with spontaneous CPV where the mass matrices
  can be constructed as linear combinations of Yukawa matrices and
  vacuum expectation values like Eq.~\eqref{eq:masslinyuk} that carry
  complex phases. This approach potentially suffers from unacceptably
  large corrections as also discussed in~\cite{Ellis:1978hq}.
\item LR-inspired models that have Eq.~\eqref{eq:argdetphases}
  automatically implemented.
\item Radiative constructions similar to~\cite{Gabrielli:2016vbb} where
  LR symmetry may not be necessary to fulfill
  condition~\eqref{eq:argdetphases}. Here we leave the field open to
  play with the ingredients.
\item Non-Abelian flavor models with Yukawa spurion fields as remaining
  vevs of heavy scalars and in such a way Eq.~\eqref{eq:argdetphases} is
  achieved dynamically by the flavon dynamics as proposed in
  Ref.~\cite{Fong:2013sba}.
\end{itemize}

As a side remark, let us note that recent investigations on minimal
left-right symmetric models hint toward the conclusion of
\(\Mat{V}_\text{CKM}^R =
\Mat{V}_\text{CKM}^L\)~\cite{Senjanovic:2014pva, Senjanovic:2015yea}.
Surprisingly, we do not get an exact equality but rather find for the
right-handed sector the angles $\theta_{12}^{\text{CKM},R} =
\theta_{12}^{\text{CKM},L}$, $\theta_{23}^{\text{CKM},R} =
\theta_{23}^{\text{CKM},L}$, and $\theta_{13}^{\text{CKM},R} \approx
\theta_{13}^{\text{CKM},L} / 10$, which results from the intricate
structure of \(\Mat{V}_\text{CKM}\) in Ref.~\cite{Hollik:2014jda}.

We do not have to rely on strict parity invariance of the fermion
Yukawa sector in order to reply the findings presented here. Parity
symmetry is broken in SM at low energies anyway and what we observe
applying the rules from above is rather a fake Parity built in the
Yukawa matrices which may be of a different origin than a GUT-inspired
remnant Parity invariant structure.

\section{Conclusions}\label{sec:concl}

We have addressed the strong \CP{} problem by following a bottom-up
approach. We have determined the necessary conditions a more
fundamental theory should have in order to intrinsically not show a
strong \CP{} phase, $\bar{\theta}=0$, not only at higher energies but
also at lower ones.  As this phase is made out of two conceptually
independent contributions, $\bar{\theta} = \theta_\text{QCD} +
\theta_\text{QFD}$, we have studied the conditions for each of them to
be zero, $\theta_\text{QCD}= \theta_\text{QFD}=0$, while
simultaneously allowing weak CPV. The first condition demands that
within a UV complete model one should have either $P$ or \CP{}
invariance. This is not a new statement as it is well known, that this
automatically sets both contributions equal to zero. However, as one
wishes to explain the observed amount of weak CPV stemming from the
quark masses, this initial symmetry must be broken. However, in
general, this induces at tree level a new strong \CP{} phase, here
denoted as $\theta_{\text{QFD}} = \arg\det\big(\Mat{M}_\uq
\Mat{M}_\dq\big) \neq 0$. The main challenge, which is naturally
present within the Nelson--Barr type of models, then basically
consists in explaining why the amount of strong CPV stemming from the
quark masses should be zero, while simultaneously a sufficiently large
value (compared to $\theta_{\text{QFD}}$) of weak CPV appears, which
is coded in the Jarlskog invariant \(J_q\) of the experimentally
measured (fitted) CKM-matrix. We have realized that there is no
difficult challenge in solving the previous problem. Splitting the
generational freedom of the gauge kinetic terms as \(\U(3) = \SU(3)
\otimes \U(1) / Z_n\), it can be clearly seen that arbitrary \(\U(1)\)
factors lead to \(\theta_{\text{QFD}} \neq 0\) while the \(\SU(3)\)
nature is responsible for \(J_q \neq 0\). Hence, the complex phases
implied by \(\theta_{\text{QFD}}\) are entirely unrelated to the
phases of weak CPV, as shown in Table \ref{table-phases}. The absence
of the strong CPV is guaranteed by imposing one of the four possible
conditions appearing in Table \ref{table-cases}. In particular, Case
II has a very minimal condition on the mass matrices such that
\(\theta_{\text{QFD}} = \sum_{q=u,d} \alpha_R^{(q)} - \alpha_L^{(q)} =
0\), if \(\alpha_L^{(q)} = \alpha_R^{(q)}\), though the basic
constraint is much weaker. (This gets important in the context of some
Grand Unification when up- and down-quark mass matrices are related to
each other.) It has been shown that minimal symmetrical requirements
on the Higgs--Yukawa interactions according
to~\cite{Saldana-Salazar:2015raa} lead to the given constraint and
non-trivial CKM-mixing. As a consequence of this, the mixing of the
right-handed sector is fixed and predicts for the right-handed
CKM-matrix $\theta_{12}^{\text{CKM},R} = \theta_{12}^{\text{CKM},L}$,
$\theta_{23}^{\text{CKM},R} = \theta_{23}^{\text{CKM},L}$, and
$\theta_{13}^{\text{CKM},R} \approx
\theta_{13}^{\text{CKM},L}/10$. This fingerprint can be tested in
future experiments within a variety of extensions of the Standard
Model.

Moreover, for the weak CPV we have showed that in the recently
proposed fermion mass ratios parametrization~\cite{Hollik:2014jda} the
leading contribution to the CKM-phase, after insertion of the value
for the mass ratios $m_\uq/m_\cq$ and $m_\dq/m_\sq$, implies the value
$\delta^{\text{q}}_{\text{CP}} \approx (1.38 \pm 0.10)\;\text{rad}$
which is in agreement to the observed one,
$\delta_{\text{CP}}^{\text{CKM}} = (1.19 \pm 0.15 )\;\text{rad}$.

We have not provided a \emph{solution} to the strong \CP{} problem but
rather argued that it can be addressed without the need of an axial
\(\U(1)\) symmetry from a flavor physics point of view by modeling the
quark mass matrices and thus does not come along with a flavored axion
(Flaxion~\cite{Ema:2016ops} or
Axiflavon~\cite{Calibbi:2016hwq}). Instead there may be several paths
to implement the condition to pass by the strong \CP{} problem via
flavor model building especially based on spontaneous breaking of the
maximal flavor group.

\section*{Acknowledgments}
The authors want to acknowledge useful conversations and a careful
reading of the manuscript to T.~Konstandin, A.~P\'erez-Lorenzana, and
M.~Spinrath. We want to thank L.L.~Salcedo for remarking the correct
decomposition of the unitary group following
Eq.~\eqref{eq:argdetphases} and W.~Buchm\"uller for an interesting
remark about a possible connection to moduli fields of string
theory. This work has been supported by CONACyT-Mexico under Contract
No.~220498. WGH acknowledges support by the DESY Fellowship Program.

During the review process, we have privately communicated with
J.-M. Gerard. We thankfully acknowledge all his remarks about our
work.

\bibliography{bib}

\providecommand{\href}[2]{#2}\begingroup\raggedright\begin{thebibliography}{10%
}\small

\bibitem{Mohapatra:1986uf}
R.~N. Mohapatra, \href{http://dx.doi.org/10.1007/978-1-4757-1928-4}{{\em
  {Unification and Supersymmetry. The Frontiers of Quark-Lepton Physics}}}.
\newblock Springer, Berlin,
1986.
\newblock
%%CITATION = INSPIRE-238200;%%.

\bibitem{Baker:2006ts}
C.~A. Baker {\em et~al.}, ``{An Improved experimental limit on the electric
  dipole moment of the neutron}'',
  \href{http://dx.doi.org/10.1103/PhysRevLett.97.131801}{{\em Phys. Rev. Lett.}
  {\bfseries 97} (2006) 131801},
\href{http://arxiv.org/abs/hep-ex/0602020}{{\ttfamily arXiv:hep-ex/0602020
  [hep-ex]}}.
%%CITATION = HEP-EX/0602020;%%.

\bibitem{Cheng:1987gp}
H.-Y. Cheng, ``{The Strong CP Problem Revisited}'',
\href{http://dx.doi.org/10.1016/0370-1573(88)90135-4}{{\em Phys. Rept.}
  {\bfseries 158} (1988) 1}.
%%CITATION = PRPLC,158,1;%%.

\bibitem{Peccei:1977hh}
R.~D. Peccei and H.~R. Quinn, ``{CP Conservation in the Presence of
  Instantons}'',
\href{http://dx.doi.org/10.1103/PhysRevLett.38.1440}{{\em Phys. Rev. Lett.}
  {\bfseries 38} (1977) 1440--1443}.
%%CITATION = PRLTA,38,1440;%%.

\bibitem{Mohapatra:1982ib}
R.~N. Mohapatra and G.~Senjanovi\'c, ``{A Simple Solution to the Strong {CP}
  Problem}'',
\href{http://dx.doi.org/10.1007/BF01407830}{{\em Z. Phys.} {\bfseries C20}
  (1983) 365}.
%%CITATION = ZEPYA,C20,365;%%.

\bibitem{Mohapatra:1983xw}
R.~N. Mohapatra, S.~Ouvry, and G.~Senjanovi\'c, ``{Automatic Solution to the
  Strong {CP} Problem in $N=1$ Supergravity Models}'',
\href{http://dx.doi.org/10.1016/0370-2693(83)90174-0}{{\em Phys. Lett.}
  {\bfseries B126} (1983) 329--333}.
%%CITATION = PHLTA,B126,329;%%.

\bibitem{Weinberg:1977ma}
S.~Weinberg, ``{A New Light Boson?}'',
\href{http://dx.doi.org/10.1103/PhysRevLett.40.223}{{\em Phys. Rev. Lett.}
  {\bfseries 40} (1978) 223--226}.
%%CITATION = PRLTA,40,223;%%.

\bibitem{Wilczek:1977pj}
F.~Wilczek, ``{Problem of Strong p and t Invariance in the Presence of
  Instantons}'',
\href{http://dx.doi.org/10.1103/PhysRevLett.40.279}{{\em Phys. Rev. Lett.}
  {\bfseries 40} (1978) 279--282}.
%%CITATION = PRLTA,40,279;%%.

\bibitem{Kuster:2008zz}
M.~Kuster, G.~Raffelt, and B.~Beltran, ``{Axions: Theory, cosmology, and
  experimental searches. Proceedings, 1st Joint ILIAS-CERN-CAST axion training,
  Geneva, Switzerland, November 30-December 2, 2005}'',
{\em Lect. Notes Phys.} {\bfseries 741} (2008) pp.1--258.
%%CITATION = LNPHA,741,pp.1;%%.

\bibitem{Ehret:2007cm}
K.~Ehret, M.~Frede, E.-A. Knabbe, D.~Kracht, A.~Lindner, N.~T. Meyer, D.~Notz,
  A.~Ringwald, and G.~Wiedemann, ``{Production and detection of axion-like
  particles in a HERA dipole magnet: Letter-of-intent for the ALPS
  experiment}'',
\href{http://arxiv.org/abs/hep-ex/0702023}{{\ttfamily arXiv:hep-ex/0702023
  [HEP-EX]}}.
%%CITATION = HEP-EX/0702023;%%.

\bibitem{Ehret:2010mh}
K.~Ehret {\em et~al.}, ``{New ALPS Results on Hidden-Sector Lightweights}'',
  \href{http://dx.doi.org/10.1016/j.physletb.2010.04.066}{{\em Phys. Lett.}
  {\bfseries B689} (2010) 149--155},
\href{http://arxiv.org/abs/1004.1313}{{\ttfamily arXiv:1004.1313 [hep-ex]}}.
%%CITATION = ARXIV:1004.1313;%%.

\bibitem{Alekhin:2015byh}
S.~Alekhin {\em et~al.}, ``{A facility to Search for Hidden Particles at the
  CERN SPS: the SHiP physics case}'',
\href{http://arxiv.org/abs/1504.04855}{{\ttfamily arXiv:1504.04855 [hep-ph]}}.
%%CITATION = ARXIV:1504.04855;%%.

\bibitem{Nelson:1983zb}
A.~E. Nelson, ``{Naturally Weak CP Violation}'',
\href{http://dx.doi.org/10.1016/0370-2693(84)92025-2}{{\em Phys. Lett.}
  {\bfseries B136} (1984) 387}.
%%CITATION = PHLTA,B136,387;%%.

\bibitem{Barr:1984qx}
S.~M. Barr, ``{Solving the Strong CP Problem Without the Peccei-Quinn
  Symmetry}'',
\href{http://dx.doi.org/10.1103/PhysRevLett.53.329}{{\em Phys. Rev. Lett.}
  {\bfseries 53} (1984) 329}.
%%CITATION = PRLTA,53,329;%%.

\bibitem{Barr:1984fh}
S.~M. Barr, ``{A Natural Class of Nonpeccei-quinn Models}'',
\href{http://dx.doi.org/10.1103/PhysRevD.30.1805}{{\em Phys. Rev.} {\bfseries
  D30} (1984) 1805}.
%%CITATION = PHRVA,D30,1805;%%.

\bibitem{Dine:2015jga}
M.~Dine and P.~Draper, ``{Challenges for the Nelson-Barr Mechanism}'',
  \href{http://dx.doi.org/10.1007/JHEP08(2015)132}{{\em JHEP} {\bfseries 08}
  (2015) 132},
\href{http://arxiv.org/abs/1506.05433}{{\ttfamily arXiv:1506.05433 [hep-ph]}}.
%%CITATION = ARXIV:1506.05433;%%.

\bibitem{Barr:1991qx}
S.~M. Barr, D.~Chang, and G.~Senjanovi\'c, ``{Strong CP problem and parity}'',
\href{http://dx.doi.org/10.1103/PhysRevLett.67.2765}{{\em Phys. Rev. Lett.}
  {\bfseries 67} (1991) 2765--2768}.
%%CITATION = PRLTA,67,2765;%%.

\bibitem{Babu:1989rb}
K.~S. Babu and R.~N. Mohapatra, ``{A Solution to the Strong {CP} Problem
  Without an Axion}'',
\href{http://dx.doi.org/10.1103/PhysRevD.41.1286}{{\em Phys. Rev.} {\bfseries
  D41} (1990) 1286}.
%%CITATION = PHRVA,D41,1286;%%.

\bibitem{Kuchimanchi:2010xs}
R.~Kuchimanchi, ``{P/CP Conserving CP/P Violation Solves Strong CP Problem}'',
  \href{http://dx.doi.org/10.1103/PhysRevD.82.116008}{{\em Phys. Rev.}
  {\bfseries D82} (2010) 116008},
\href{http://arxiv.org/abs/1009.5961}{{\ttfamily arXiv:1009.5961 [hep-ph]}}.
%%CITATION = ARXIV:1009.5961;%%.

\bibitem{Kuchimanchi:2012xb}
R.~Kuchimanchi, ``{Maximal CP and Bounds on the Neutron Electric Dipole Moment
  from P and CP Breaking}'',
  \href{http://dx.doi.org/10.1103/PhysRevD.86.036002}{{\em Phys. Rev.}
  {\bfseries D86} (2012) 036002},
\href{http://arxiv.org/abs/1203.2772}{{\ttfamily arXiv:1203.2772 [hep-ph]}}.
%%CITATION = ARXIV:1203.2772;%%.

\bibitem{Georgi:1978xz}
H.~Georgi, ``{A Model of Soft CP Violation}'',
{\em Hadronic J.} {\bfseries 1} (1978) 155.
%%CITATION = HADJM,1,155;%%.

\bibitem{Antusch:2013rla}
S.~Antusch, M.~Holthausen, M.~A. Schmidt, and M.~Spinrath, ``{Solving the
  Strong CP Problem with Discrete Symmetries and the Right Unitarity
  Triangle}'', \href{http://dx.doi.org/10.1016/j.nuclphysb.2013.10.028}{{\em
  Nucl. Phys.} {\bfseries B877} (2013) 752--771},
\href{http://arxiv.org/abs/1307.0710}{{\ttfamily arXiv:1307.0710 [hep-ph]}}.
%%CITATION = ARXIV:1307.0710;%%.

\bibitem{Maiezza:2014ala}
A.~Maiezza and M.~Nemev\v{s}ek, ``{Strong P invariance, neutron electric dipole
  moment, and minimal left-right parity at LHC}'',
  \href{http://dx.doi.org/10.1103/PhysRevD.90.095002}{{\em Phys. Rev.}
  {\bfseries D90} no.~9, (2014) 095002},
\href{http://arxiv.org/abs/1407.3678}{{\ttfamily arXiv:1407.3678 [hep-ph]}}.
%%CITATION = ARXIV:1407.3678;%%.

\bibitem{Senjanovic:2014pva}
G.~Senjanovi\'c and V.~Tello, ``{Right Handed Quark Mixing in Left-Right
  Symmetric Theory}'',
  \href{http://dx.doi.org/10.1103/PhysRevLett.114.071801}{{\em Phys. Rev.
  Lett.} {\bfseries 114} no.~7, (2015) 071801},
\href{http://arxiv.org/abs/1408.3835}{{\ttfamily arXiv:1408.3835 [hep-ph]}}.
%%CITATION = ARXIV:1408.3835;%%.

\bibitem{Senjanovic:2015yea}
G.~Senjanovi\'c and V.~Tello, ``{Restoration of Parity and the Right-Handed
  Analog of the CKM Matrix}'',
\href{http://arxiv.org/abs/1502.05704}{{\ttfamily arXiv:1502.05704 [hep-ph]}}.
%%CITATION = ARXIV:1502.05704;%%.

\bibitem{Barr:1996wx}
S.~M. Barr, ``{Supersymmetric solutions to the strong CP problem}'',
  \href{http://dx.doi.org/10.1103/PhysRevD.56.1475}{{\em Phys. Rev.} {\bfseries
  D56} (1997) 1475--1480},
\href{http://arxiv.org/abs/hep-ph/9612396}{{\ttfamily arXiv:hep-ph/9612396
  [hep-ph]}}.
%%CITATION = HEP-PH/9612396;%%.

\bibitem{Hiller:2001qg}
G.~Hiller and M.~Schmaltz, ``{Solving the strong CP problem with
  supersymmetry}'', \href{http://dx.doi.org/10.1016/S0370-2693(01)00814-0}{{\em
  Phys. Lett.} {\bfseries B514} (2001) 263--268},
\href{http://arxiv.org/abs/hep-ph/0105254}{{\ttfamily arXiv:hep-ph/0105254
  [hep-ph]}}.
%%CITATION = HEP-PH/0105254;%%.

\bibitem{Ellis:1978hq}
J.~R. Ellis and M.~K. Gaillard, ``{Strong and Weak CP Violation}'',
\href{http://dx.doi.org/10.1016/0550-3213(79)90297-9}{{\em Nucl. Phys.}
  {\bfseries B150} (1979) 141--162}.
%%CITATION = NUPHA,B150,141;%%.

\bibitem{Konstandin}
 Private communication with W.~Buchmuller.

\bibitem{Hollik:2014jda}
W.~G. Hollik and U.~J. Salda\~na Salazar, ``{The double mass hierarchy pattern:
  simultaneously understanding quark and lepton mixing}'',
  \href{http://dx.doi.org/10.1016/j.nuclphysb.2015.01.019}{{\em Nucl. Phys.}
  {\bfseries B892} (2015) 364--389},
\href{http://arxiv.org/abs/1411.3549}{{\ttfamily arXiv:1411.3549 [hep-ph]}}.
%%CITATION = ARXIV:1411.3549;%%.

\bibitem{Gerard}
 Private communication with J.-M. Gerard.

\bibitem{Gerard:2012ft}
J.-M. Gerard and Z.-z. Xing, ``{Flavor Mixing Democracy and Minimal CP
  Violation}'', \href{http://dx.doi.org/10.1016/j.physletb.2012.05.037}{{\em
  Phys. Lett.} {\bfseries B713} (2012) 29--34},
\href{http://arxiv.org/abs/1203.0496}{{\ttfamily arXiv:1203.0496 [hep-ph]}}.
%%CITATION = ARXIV:1203.0496;%%.

\bibitem{Jarlskog:1985ht}
C.~Jarlskog, ``{Commutator of the Quark Mass Matrices in the Standard
  Electroweak Model and a Measure of Maximal CP Violation}'',
\href{http://dx.doi.org/10.1103/PhysRevLett.55.1039}{{\em Phys. Rev. Lett.}
  {\bfseries 55} (1985) 1039}.
%%CITATION = PRLTA,55,1039;%%.

\bibitem{Gerard:2008nc}
J.-M. Gerard, ``{Mass Issues in Fundamental Interactions}'', in {\em {2008
  European School of High-Energy Physics, Herbeumont-sur-Semois, Belgium, 8-21
  June 2008}}, pp.~281--314.
\newblock 2008.
\newblock \href{http://arxiv.org/abs/0811.0540}{{\ttfamily arXiv:0811.0540
  [hep-ph]}}.
\newblock
\url{http://inspirehep.net/record/801581/files/arXiv:0811.0540.pdf}.
\newblock
%%CITATION = ARXIV:0811.0540;%%.

\bibitem{Aoki:2013ldr}
S.~Aoki {\em et~al.}, ``{Review of lattice results concerning low-energy
  particle physics}'',
  \href{http://dx.doi.org/10.1140/epjc/s10052-014-2890-7}{{\em Eur. Phys. J.}
  {\bfseries C74} (2014) 2890},
\href{http://arxiv.org/abs/1310.8555}{{\ttfamily arXiv:1310.8555 [hep-lat]}}.
%%CITATION = ARXIV:1310.8555;%%.

\bibitem{Salcedo}
 Private communication with L. L. Salcedo.

\bibitem{Agashe:2014kda}
{\bfseries Particle Data Group} , K.~A. Olive {\em et~al.}, ``{Review of
  Particle Physics}'',
\href{http://dx.doi.org/10.1088/1674-1137/38/9/090001}{{\em Chin. Phys.}
  {\bfseries C38} (2014) 090001}.
%%CITATION = CHPHD,C38,090001;%%.

\bibitem{Grimus:2010ak}
W.~Grimus and P.~O. Ludl, ``{Principal series of finite subgroups of SU(3)}'',
  \href{http://dx.doi.org/10.1088/1751-8113/43/44/445209}{{\em J. Phys.}
  {\bfseries A43} (2010) 445209},
\href{http://arxiv.org/abs/1006.0098}{{\ttfamily arXiv:1006.0098 [hep-ph]}}.
%%CITATION = ARXIV:1006.0098;%%.

\bibitem{Grimus:2011fk}
W.~Grimus and P.~O. Ludl, ``{Finite flavour groups of fermions}'',
  \href{http://dx.doi.org/10.1088/1751-8113/45/23/233001}{{\em J. Phys.}
  {\bfseries A45} (2012) 233001},
\href{http://arxiv.org/abs/1110.6376}{{\ttfamily arXiv:1110.6376 [hep-ph]}}.
%%CITATION = ARXIV:1110.6376;%%.

\bibitem{Holdom:1999ny}
B.~Holdom, ``{Nonstandard order parameters and the origin of CP violation}'',
  \href{http://dx.doi.org/10.1103/PhysRevD.61.011702}{{\em Phys. Rev.}
  {\bfseries D61} (2000) 011702},
\href{http://arxiv.org/abs/hep-ph/9907361}{{\ttfamily arXiv:hep-ph/9907361
  [hep-ph]}}.
%%CITATION = HEP-PH/9907361;%%.

\bibitem{Gatto:1968ss}
R.~Gatto, G.~Sartori, and M.~Tonin, ``{Weak Selfmasses, Cabibbo Angle, and
  Broken SU(2) x SU(2)}'',
\href{http://dx.doi.org/10.1016/0370-2693(68)90150-0}{{\em Phys. Lett.}
  {\bfseries B28} (1968) 128--130}.
%%CITATION = PHLTA,B28,128;%%.

\bibitem{Saldana-Salazar:2015raa}
U.~Salda\~na Salazar, ``{The flavor-blind principle: A symmetrical approach to
  the Gatto-Sartori-Tonin relation}'',
  \href{http://dx.doi.org/10.1103/PhysRevD.93.013002}{{\em Phys. Rev.}
  {\bfseries D93} no.~1, (2016) 013002},
\href{http://arxiv.org/abs/1509.08877}{{\ttfamily arXiv:1509.08877 [hep-ph]}}.
%%CITATION = ARXIV:1509.08877;%%.

\bibitem{Antusch:2009hq}
S.~Antusch, S.~F. King, M.~Malinsky, and M.~Spinrath, ``{Quark mixing sum rules
  and the right unitarity triangle}'',
  \href{http://dx.doi.org/10.1103/PhysRevD.81.033008}{{\em Phys. Rev.}
  {\bfseries D81} (2010) 033008},
\href{http://arxiv.org/abs/0910.5127}{{\ttfamily arXiv:0910.5127 [hep-ph]}}.
%%CITATION = ARXIV:0910.5127;%%.

\bibitem{Mohapatra:1997su}
R.~N. Mohapatra, A.~Rasin, and G.~Senjanovic, ``{P, C and strong CP in
  left-right supersymmetric models}'',
  \href{http://dx.doi.org/10.1103/PhysRevLett.79.4744}{{\em Phys. Rev. Lett.}
  {\bfseries 79} (1997) 4744--4747},
\href{http://arxiv.org/abs/hep-ph/9707281}{{\ttfamily arXiv:hep-ph/9707281
  [hep-ph]}}.
%%CITATION = HEP-PH/9707281;%%.

\bibitem{Kuchimanchi:1995rp}
R.~Kuchimanchi, ``{Solution to the strong CP problem: Supersymmetry with
  parity}'', \href{http://dx.doi.org/10.1103/PhysRevLett.76.3486}{{\em Phys.
  Rev. Lett.} {\bfseries 76} (1996) 3486--3489},
\href{http://arxiv.org/abs/hep-ph/9511376}{{\ttfamily arXiv:hep-ph/9511376
  [hep-ph]}}.
%%CITATION = HEP-PH/9511376;%%.

\bibitem{Chang:1993eq}
S.~Chang and J.~E. Kim, ``{Fermion doubling and a natural solution of the
  strong CP problem}'', \href{http://dx.doi.org/10.1103/PhysRevD.50.2218}{{\em
  Phys. Rev.} {\bfseries D50} (1994) 2218--2224},
\href{http://arxiv.org/abs/hep-ph/9309274}{{\ttfamily arXiv:hep-ph/9309274
  [hep-ph]}}.
%%CITATION = HEP-PH/9309274;%%.

\bibitem{Gabrielli:2016vbb}
E.~Gabrielli, L.~Marzola, and M.~Raidal, ``{Radiative Yukawa Couplings in the
  Simplest Left-Right Symmetric Model}'',
\href{http://arxiv.org/abs/1611.00009}{{\ttfamily arXiv:1611.00009 [hep-ph]}}.
%%CITATION = ARXIV:1611.00009;%%.

\bibitem{Ema:2016ops}
Y.~Ema, K.~Hamaguchi, T.~Moroi, and K.~Nakayama, ``{Flaxion: a minimal
  extension to solve puzzles in the standard model}'',
\href{http://arxiv.org/abs/1612.05492}{{\ttfamily arXiv:1612.05492 [hep-ph]}}.
%%CITATION = ARXIV:1612.05492;%%.

\bibitem{Calibbi:2016hwq}
L.~Calibbi, F.~Goertz, D.~Redigolo, R.~Ziegler, and J.~Zupan, ``{The
  Axiflavon}'',
\href{http://arxiv.org/abs/1612.08040}{{\ttfamily arXiv:1612.08040 [hep-ph]}}.
%%CITATION = ARXIV:1612.08040;%%.

\bibitem{Fong:2013sba}
C.~S. Fong and E.~Nardi, ``{Spontaneous Breaking of Flavor Symmetry Avoids the
  Strong CP Problem}'',
  \href{http://dx.doi.org/10.1103/PhysRevLett.111.061601}{{\em Phys. Rev.
  Lett.} {\bfseries 111} no.~6, (2013) 061601},
\href{http://arxiv.org/abs/1305.1627}{{\ttfamily arXiv:1305.1627 [hep-ph]}}.
%%CITATION = ARXIV:1305.1627;%%.

\end{thebibliography}\endgroup
\end{document}